\documentclass[aps,prl,reprint,groupedaddress,twocolumn]{revtex4-1}

\usepackage{xcolor}
\usepackage[utf8]{inputenc}
\usepackage{graphicx}
\usepackage{amsmath}
\usepackage{amssymb}

\newcommand{\ket}[1]{\big|#1\big>}

\newcommand{\bk}{\mathbf{k}}

\newcommand{\bq}{\mathbf{q}}

\def\dz2{d$_{\text{z}^2}$}

\def\dx2y2{d$_{\text{x}^2\text{y}^2}$}
\def\G0W0{G$_0$W$_0$}
\def\scGW0{scGW$_0$}

\begin{document}


\title{Frequency-dependent substrate screening of excitons in atomically thin transition metal dichalcogenide semiconductors}


\author{A. Steinhoff}
\affiliation{Institut f\"ur Theoretische Physik, Universit\"at Bremen, P.O. Box 330 440, 28334 Bremen, Germany}

\author{T.O. Wehling}
\affiliation{Institut f\"ur Theoretische Physik, Universit\"at Bremen, P.O. Box 330 440, 28334 Bremen, Germany}
\affiliation{Bremen Center for Computational Materials Science, Universit\"at Bremen, 28334 Bremen, Germany}
\affiliation{MAPEX Center for Materials and Processes, Universit\"at Bremen, 28359 Bremen, Germany}

\author{M. R\"osner}

\affiliation{Department of Physics and Astronomy, University of Southern California, 825 Bloom Walk, ACB 439, Los Angeles, CA 90089-0484, USA}




\begin{abstract}

Atomically thin layers of transition metal dichalcogenides (TMDCs) exhibit exceptionally strong Coulomb interaction between charge carriers due to the two-dimensional carrier confinement in connection with weak dielectric screening. The van der Waals nature of interlayer coupling makes it easy to integrate TMDC layers into heterostructures with different dielectric or metallic substrates. This allows to tailor electronic and optical properties of these materials, as Coulomb interaction inside atomically thin layers is very susceptible to screening by the environment. 
Here we theoretically investigate dynamical screening effects in TMDCs due to bulk substrates doped with carriers over a large density range, thereby offering three-dimensional plasmons as tunable degree of freedom. We report a wide compensation of renormalization effects leading to a spectrally more stable exciton than predicted for static substrate screening, even if plasmons and excitons are in resonance. We also find a nontrivial dependence of the single-particle band gap on substrate doping density due to dynamical screening. Our investigation provides microscopic insight into the mechanisms that allow for manipulations of TMDC excitons by means of arbitrary plasmonic environments on the nanoscale.

\end{abstract}


\maketitle

\section{Introduction}

Monolayers of transition metal dichalcogenide (TMDC) semiconductors are well-suited as active materials in optoelectronic devices such as light-emitting diodes \cite{pospischil_solar-energy_2014, baugher_optoelectronic_2014, ross_electrically_2014, withers_light-emitting_2015}, solar cells \cite{pospischil_solar-energy_2014, baugher_optoelectronic_2014}, and lasers \cite{wu_monolayer_2015, ye_monolayer_2015, salehzadeh_optically_2015, li_room-temperature_2017}. A central aspect of these applications is the combinability with different substrates or other two-dimensional materials in functional van der Waals-heterostructures. \cite{geim_van_2013} 
Here, fascinating prospects arise from the possibility to engineer electronic and optical properties by manipulating the Coulomb interaction in atomically thin materials via its dielectric environment 
\cite{berkelbach_theory_2013, latini_excitons_2015, steinke_non-invasive_2017, trolle_model_2017, raja_coulomb_2017, steinhoff_exciton_2017,meckbach_influence_2018,florian_dielectric_2018,meckbach_giant_2018}.

Most of the available theoretical approaches to describe these effects have in common that environmental screening is treated by means of a macroscopic model for the dielectric function of the heterostructure formed by an active TMDC layer and its environment. Here, the environment is often described by a static dielectric function. Besides these environmental or substrate screening effects, screening due to free or bound charge carriers in the TMDC layer itself has been considered. Excited electron-hole pairs in the TMDC layer have been shown to reduce single-particle band gaps and exciton binding energies \cite{steinhoff_influence_2014, pogna_photo-induced_2016, sie_observation_2017, steinhoff_exciton_2017, cunningham_photoinduced_2017}. Also, dynamical and thus frequency-dependent screening effects due to doped charge carriers have been investigated, yielding spectral shifts of excitons \cite{gao_dynamical_2016}, inter- and intra-valley plamonics \cite{groenewald_valley_2016, dery_theory_2016} and optical sidebands induced by exciton-plasmon coupling \cite{van_tuan_marrying_2017, scharf_dynamical_2018}. 

If we take dynamical screening effects into account, it is important to realize that band gaps and excitons are sensitive to screening at different frequencies: While the band gap turns out to be rather affected by low frequencies and thus by low-energy plasmons, the characteristic energy scale of excitons is their binding energy of several hundred meV. It is a teasing but still open question what happens if the exciton binding energy matches a resonance in the substrate dielectric function. It has been demonstrated recently that polaritons emerge from the strong coupling of monolayer WS$_2$ excitons to plasmons in a gold substrate \cite{goncalves_plasmon-exciton_2018}. While polaritons are usually associated with the transverse dielectric response of a medium at optical frequencies \cite{haug_electron_1984}, we focus here on effects due to longitudinal excitations at lower frequencies in the THz range.

In this paper, we investigate dynamical screening effects on excitons and the single-particle band gap in a TMDC monolayer. Therefore, we consider a heterostructure formed by the monolayer on top of a metallic bulk substrate. The latter hosts three-dimensional plasmons which can be tuned by the substrate doping level, thereby yielding a variable dynamical screening environment. We study the resulting effects by solving a Bethe-Salpeter equation (BSE) in dynamically screened ladder approximation using non-equilibrium Green functions. To account for the dynamical screening effects of the environment, we include the substrate dielectric function via a macroscopic linearized Keldysh model. We find the exciton to be spectrally stable over a wide substrate doping range (and thus a wide range of substrate plasmon frequencies) before showing a red shift of up to several ten meV. The red shift is systematically smaller than estimated by a theory based on static screening. At the same time, a nontrivial dependence of band-gap shrinkage on substrate doping density is found, where the band-gap reduction  is again overestimated in the static case. No specific exciton-plasmon resonance is observed at room temperature due to the low thermal-equilibrium population of high-energy plasmons and efficient compensation between different renormalization effects. Only at elevated temperatures, a resonance emerges in the spectral position of the exciton.

Our results open an avenue for a microscopic understanding of the TMDC exciton manipulation by means of complex plasmonic nanostructures. 


%
%

\section{Dielectric function including dynamical substrate screening}
\begin{figure}[h!t]
\centering
\includegraphics[width=.7\columnwidth]{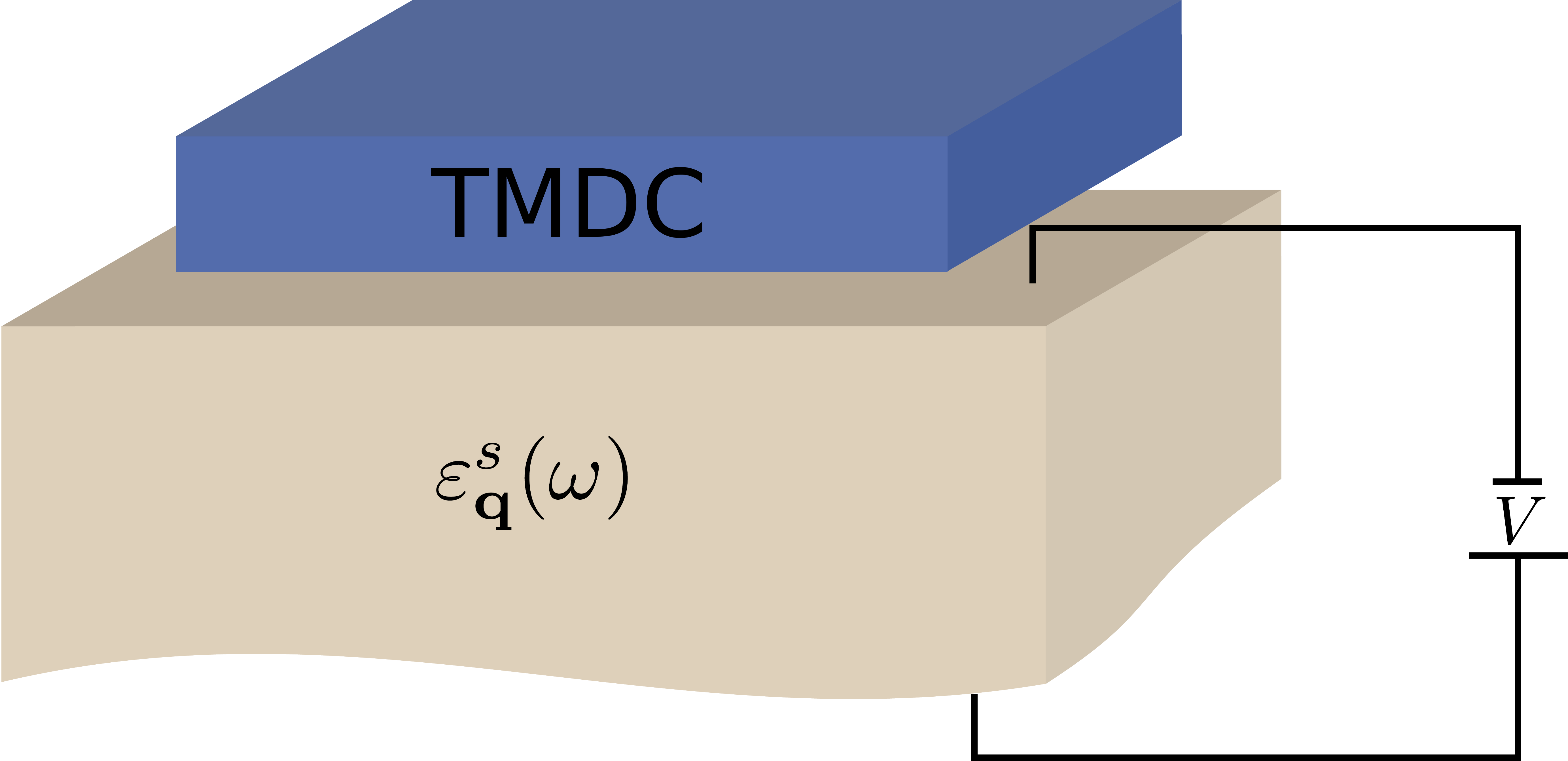}
\caption{Schematic of the heterostructure composed of a single TMDC layer 
placed on a three-dimensional substrate with dielectric function $\varepsilon^{\textrm{s}}_{\bq}(\omega)$. The substrate is assumed to be electrically contacted to tune its dynamical dielectric properties.}
\label{fig:scheme}
\end{figure}
Our goal is to describe dynamical screening effects on charge carriers in a TMDC layer due to plasmonic excitations in a substrate as shown in Fig.~\ref{fig:scheme}. To this end, we derive a model for the macroscopic screened carrier-carrier interaction in the TMDC embedded in the heterostructure (HS) $W^{\textrm{HS}}_{\bq}(\omega)$. It is given by the bare Coulomb potential $V_{\bq}$ divided by the macroscopic longitudinal dielectric function of the heterostructure $\varepsilon^{\textrm{HS}}_{\bq}(\omega)$:
\begin{equation}
 \begin{split}
 W^{\textrm{HS}}_{\bq}(\omega)=V_{\bq}\varepsilon^{\textrm{HS},-1}_{\bq}(\omega)\,.
 \end{split}
\label{eq:carrier_carrier}
\end{equation}
Via the screened potential $W^{\textrm{HS}}$ the TMDC charge carriers and, in particular, excitons couple to excitations of the substrate material that are described by the substrate dielectric function $\varepsilon^{\textrm{s}}_{\bq}(\omega)$, which is part of $\varepsilon^{\textrm{HS}}_{\bq}(\omega)$.
The simplest approximation to the interaction potential in a two-dimensional layer surrounded by dielectric material is a linearized Keldysh potential \cite{keldysh_coulomb_1979, berkelbach_theory_2013} corresponding to the heterostructure dielectric function 
\begin{equation}
 \begin{split}
 \varepsilon^{\textrm{HS}}_{\bq}(\omega)=\kappa_{\bq}(\omega)+r_0 q
 \end{split}
\label{eq:Keldysh_DF}
\end{equation}
with $\kappa_{\bq}(\omega)= \frac{1+\varepsilon^{\textrm{s}}_{\bq}(\omega)}{2}$ and $r_0=\frac{\chi_{2\textrm{d}}}{2\varepsilon_0}$. Here $\chi_{2\textrm{d}}$ is the polarizability of the TMDC layer that can be obtained from first-principle calculations. It accounts for background screening due to electrons in the filled valence bands. The linearized Keldysh potential constitutes the long-wavelength limit of the full interaction potential, which has a more complex momentum dependence \cite{latini_excitons_2015, rosner_wannier_2015}. This approximation is justified as we focus on the quasi-particle band edge and excitons in the K valley, which are localized around the valley minimum \cite{qiu_optical_2013, steinhoff_influence_2014}. Moreover, the internal TMDC susceptibility is in principle frequency-dependent. This becomes particularly important at frequencies on the order of the quasi-particle band gap, where inter-band polarizations are possible. As a detailed analysis will show later on, the quasi-particle gap energy is far above the energy scale we consider here. Renormalizations of both, excitons and the gap energy itself are sensitive to screening at lower frequencies. 
\par
The substrate dielectric function in principle accounts for all possible longitudinal excitations that may couple to the Coulomb interaction potential between charges in the TMDC. In the case of optical phonons or plasmons the frequency dependence of the dielectric function is captured by a Lorentz-oscillator model \cite{haug_electron_1984, karimi_dielectric_2016}, where the parameters can be either adjusted to fit experimental data or calculated from first principles.
As we will later focus on plasmonic excitations of the substrate, we specify the dielectric function in the following.
The plasmons are described by a single plasmon pole (SPP) model \cite{lundqvist_single-particle_1967, haug_electron_1984}:
\begin{equation}
 \begin{split}
 \varepsilon^{\textrm{s}}_{\bq}(\omega)=\varepsilon_{\infty}\varepsilon^{\textrm{SPP}}_{\bq}(\omega)= \varepsilon_{\infty}\Big(1+\frac{\Omega_{\textrm{pl}}^2}{\tilde{\omega}_{\bq}^2-\Omega_{\textrm{pl}}^2-\omega^2-i\gamma\omega}\Big)\,
 \end{split}
\label{eq:substrate_DF_plasmon}
\end{equation}
with the plasma frequency $\Omega_{\textrm{pl}}=(\frac{e^2 n}{\varepsilon_0\varepsilon_{\infty}m})^{1/2}$ and the plasmon pole frequency $\tilde{\omega}_{\bq}= (\Omega_{\textrm{pl}}^2(1+\frac{q^2}{\kappa^2})+v_{\bq}^2)^{1/2}$.
Here, $n$ is the density of carriers in the substrate, $m$ is the substrate carrier mass and $\kappa=(\frac{e^2}{\pi^2\varepsilon_0\varepsilon_{\infty}\hbar^2}k_{\textrm{F}}m)^{1/2}$ with $k_{\textrm{F}}=(3\pi^2 n)^{1/3}$ is the inverse screening length in Thomas-Fermi ($T=0$) approximation. $\kappa$ describes metallic screening in the static and long-wavelength limit of the dielectric function and $v_{\bq}=\frac{\hbar^2 q^2}{2 m}$ is introduced to account for the pair continuum \cite{lundqvist_single-particle_1967}. 
Lastly, $\varepsilon_{\infty}$ is the high-frequency limit of the dielectric function due to electrons in inner shells of the substrate atoms that provide a static background screening.
Combining Eqs.~(\ref{eq:Keldysh_DF}) and (\ref{eq:substrate_DF_plasmon}) we obtain the total dielectric function
\begin{equation}
 \begin{split}
 \varepsilon^{\textrm{HS}}_{\bq}(\omega)=\tilde{\varepsilon}_{\bq}+\frac{s^2}{\omega_{0,\bq}^2-\omega^2-i\gamma\omega}\,,
 \end{split}
\label{eq:total_DF}
\end{equation}
where we introduced $\tilde{\varepsilon}_{\bq}= \frac{1+\varepsilon_{\infty}}{2}+\frac{\chi_{2\textrm{d}}}{2\varepsilon_0} q$, $s^2=\frac{\varepsilon_{\infty}\Omega_{\textrm{pl}}^2}{2}$ and $\omega_{0,\bq}^2= \Omega_{\textrm{pl}}^2\frac{q^2}{\kappa^2}+v_{\bq}^2$. The inverse dielectric function is then given by
\begin{equation}
 \begin{split}
 \varepsilon^{\textrm{HS},-1}_{\bq}(\omega) & = \frac{\omega_{0,\bq}^2-\omega^2-i\gamma\omega}{\tilde{\varepsilon}_{\bq}(\omega_{0,\bq}^2-\omega^2-i\gamma\omega)+s^2}  \\ 
 & =\tilde{\varepsilon}^{-1}_{\bq}-\frac{s^2/\tilde{\varepsilon}_{\bq}}{\tilde{\varepsilon}_{\bq}(\omega_{0,\bq}^2-\omega^2-i\gamma\omega)+s^2} \,.
 \end{split}
\label{eq:total_invDF}
\end{equation}
It separates into a static purely electronic part $\tilde{\varepsilon}^{-1}_{\bq}$ due to inner-shell electrons in substrate and TMDC layer and a dynamical part due to bosonic excitations. This separation has been discussed more generally using a mean-field treatment of the dielectric response \cite{allen_total_1988} or nonequilibrium Green functions \cite{leeuwen_first-principles_2004}. For later purposes we calculate the imaginary part of the inverse dielectric function, which is also known as the loss function:
\begin{equation}
 \begin{split}
 \textrm{Im}\,\varepsilon^{\textrm{HS},-1}_{\bq}(\omega) = -\frac{\gamma\omega s^2}{(\tilde{\varepsilon}_{\bq}(\omega_{0,\bq}^2-\omega^2))^2+\tilde{\varepsilon}_{\bq}^2\gamma^2\omega^2}\,.
 \end{split}
\label{eq:im_total_invDF}
\end{equation}
The loss function exhibits two resonances at 
\begin{equation}
 \begin{split}
\omega_{1/2,\bq}=\pm\sqrt{\omega_{0,\bq}^2+s^2/\tilde{\varepsilon}_{\bq}}\,, 
 \end{split}
\label{eq:resonance}
\end{equation}
which are shifted from the original resonance frequency of the dielectric function. Further, the resonances are modified by electronic background screening via $\tilde{\varepsilon}_{\bq}$. 
In the limit of infinite excitation lifetimes or vanishing $\gamma$, the loss function can be expressed by a sum of Delta distributions using the property $\delta(f(x))=\sum_i \delta(x-x_i)/|f'(x)|_{x=x_i}$ for $f(x_i)=0$:
\begin{equation}
 \begin{split}
 \textrm{Im}\,\varepsilon^{\textrm{HS},-1}_{\bq}(\omega) = -\frac{s^2}{\tilde{\varepsilon}_{\bq}^2}\frac{\pi}{2\omega_{1,\bq}}(\delta(\omega-\omega_{1,\bq})-\delta(\omega+\omega_{1,\bq}))\,.
 \end{split}
\label{eq:im_total_invDF_delta}
\end{equation}
This expression shows that the electron-boson interaction given by the loss function is screened by the inner electrons in substrate and TMDC layer described by $\tilde{\varepsilon}_{\bq}^2$. The squared electron dielectric function can be understood from the fact that the electron-boson interaction scales with the square of the matrix element $s$.\cite{haug_electron_1984, allen_total_1988, schafer_semiconductor_2002, leeuwen_first-principles_2004} 
\par
As a next step, we link the bosonic excitations of the substrate to the electronic properties of the TMDC layer by setting up a BSE including carrier-carrier interaction that is screened by the heterostructure dielectric function $\varepsilon^{\textrm{HS}}_{\bq}(\omega)$.

\section{Bethe-Salpeter equation for heterostructures}

The BSE describes the two-particle spectrum under the influence of carrier-carrier interaction. To include excitons as bound two-particle states and the effect of screening at the same time, carrier-carrier interaction has to be treated in screened-ladder approximation. \cite{kremp_quantum_2005} Quasi-particle energy renormalizations are then included consistently on a GW level. The BSE is often used in statically screened ladder approximation for first-principle calculations of exciton spectra \cite{marini_yambo:_2009, qiu_optical_2013}. Recently, it has also been used in dynamically screened ladder approximation to describe effects of exciton-plasmon coupling in TMDCs \cite{gao_dynamical_2016, van_tuan_marrying_2017, scharf_dynamical_2018}. In these references, either the formalism of zero-temperature or Matsubara Green functions is applied. 
\par
Here we apply the technique of nonequilibrium or Schwinger-Keldysh Green functions, which has the advantage that equilibrium states at finite temperatures can be treated directly in the domain of real frequencies as long as initial correlations are negligible \cite{kremp_quantum_2005}. Moreover, a closed equation for two-particle Green functions in screened-ladder approximation involving only one frequency can be derived as shown in \citenum{bornath_two-particle_1999}. To this end, the semi-group property of free single-particle propagators is used to rearrange the integral kernel, providing the BSE with the same algebraic structure as the single-particle Dyson-Keldysh equation.
We present the BSE in general form in the Appendix, where we also derive a particular version that can be applied to the heterostructure shown in Fig.~\ref{fig:scheme}.
In the electron-hole (e-h) picture, assuming one electron and hole band, respectively, and in the absence of electron-hole pair excitation, the BSE in screened-ladder approximation for heterostructures is given by
\begin{equation}
 \begin{split}
  & (\varepsilon_{\bk}^{\textrm{G}_0\textrm{W}_0,\textrm{e}}+\varepsilon_{\bk}^{\textrm{G}_0\textrm{W}_0,\textrm{h}}+\Sigma_{\bk}^{\textrm{SXCH,e}}+\Sigma_{\bk}^{\textrm{SXCH,h}}    \\&+\Delta^{\textrm{dyn,e}}_{\bk}(\omega)+\Delta^{\textrm{dyn,h}}_{\bk}(\omega)-\hbar\omega)\psi^{\textrm{eh}}_{\bk}(\omega) \\
  &-\frac{1}{\mathcal{A}}\sum_{\bk'}W_{\bk\bk'}^{\textrm{stat,eh}}\psi^{\textrm{eh}}_{\bk'}(\omega)\\&-\frac{1}{\mathcal{A}}\sum_{\bk'}\Xi^{\textrm{dyn,eh}}_{\bk\bk'}(\omega)\psi^{\textrm{eh}}_{\bk'}(\omega)=0\,.
 \end{split}
\label{eq:BSE_HS}
\end{equation}
Here, $\psi^{\textrm{eh}}_{\bk}(\omega)$ denotes the two-particle spectral function, $\varepsilon_{\bk}^{\textrm{G}_0\textrm{W}_0,\textrm{e/h}}$ are electron and hole energies containing renormalizations on a G$_0$W$_0$-level from the freestanding layer and $\mathcal{A}$ is the area of the TMDC layer.
As the dielectric function of the heterostructure contains static (frequency-independent) as well as dynamical (frequency-dependent) screening effects, there are contributions due to statically and dynamically screened carrier-carrier interaction to the BSE. Dynamical screening contributions 
are contained in the so-called diagonal and off-diagonal correlations $\Delta^{\textrm{dyn,e/h}}_{\bk}(\omega)$ and $ \Xi^{\textrm{dyn,eh}}_{\bk\bk'}(\omega)$. They essentially describe renormalizations due to substrate-plasmon-assisted scattering processes of TMDC carriers.
On the other hand, as we discussed in the previous section, there are background contributions to screening from inner-shell electrons in the substrate and TMDC layer that we assume to be frequency-independent. These contributions enter the BSE via the statically screened electron-hole interaction 
\begin{equation}
 \begin{split}
W_{\bk\bk'}^{\textrm{stat,eh}}=V^{\textrm{eh}}_{\bk\bk'}\tilde{\varepsilon}^{-1}_{\bk-\bk'}\,
 \end{split}
\label{eq:static_e_h_term}
\end{equation}
with the bare electron-hole interaction 
\begin{equation}
 \begin{split}
&V^{\textrm{eh}}_{\bk\bk'}=\\ &\int\int\textbf{dr}^3\textbf{dr}'^3(\Phi_{\bk}^{\textrm{e}}(\textbf{r}))^*(\Phi^{\textrm{h}}_{\bk'}(\textbf{r}'))^*\frac{1}{|\textbf{r}-\textbf{r}'|}\Phi^{\textrm{h}}_{\bk}(\textbf{r}')\Phi^{\textrm{e}}_{\bk'}(\textbf{r})\,.
 \end{split}
\label{eq:bare_Coul}
\end{equation}
Further static contributions are given by screened-exchange (SX) as well as Coulomb-hole (CH) self-energies $\Sigma_{\bk}^{\textrm{SXCH,e/h}}$ \cite{haug_quantum_2004, steinhoff_influence_2014} describing frequency-independent band-structure renormalizations. 
Eq.~(\ref{eq:BSE_HS}) constitutes a frequency-dependent eigenvalue problem in the space of momentum states $\ket{\bk}$ that yields the eigenvalues $E_{\alpha}(\omega)$ and eigenstates $\ket{\psi_{\alpha}(\omega)}$ describing the two-particle spectrum of the TMDC monolayer on a substrate. Due to the frequency dependence of the correlations, the eigenvalues of bound excitons $E_{\textrm{X}}$ and of the quasi-particle band gap $E_{\textrm{Gap}}$ are subject to different renormalization effects. For example, $E_{\textrm{X}}$ is obtained by diagonalizing the BSE at $\hbar\omega=E_{\textrm{X}}$, which makes it a self-consistency problem. 
\par
We apply an effective-mass approximation to the TMDC band structure and introduce the linearized Keldysh potential as given by Eqs.~(\ref{eq:carrier_carrier}) and (\ref{eq:total_invDF}) for the screened carrier-carrier interaction matrix elements entering the correlation integrals,
\begin{equation}
 \begin{split}
W^{\textrm{HS,ee}}_{\bk\bk'}(\omega)=W^{\textrm{HS,hh}}_{\bk\bk'}(\omega)=W^{\textrm{HS,eh}}_{\bk\bk'}(\omega)=W^{\textrm{HS}}_{\bk-\bk'}(\omega)\,,
 \end{split}
\label{eq:replace_coulomb}
\end{equation}
to explicitly evaluate all terms of the BSE. The bare carrier-carrier interaction is described by the two-dimensional Coulomb potential $V_{\bq}=\frac{e^2}{2\varepsilon_0|\bq|}$. The hole and electron dispersion are given by 
\begin{equation}
 \begin{split}
&\varepsilon_{\bk}^{\textrm{G}_0\textrm{W}_0,\textrm{h}}=\frac{\hbar^2 k^2}{2 m_{\textrm{h}}}=\alpha_{\textrm{h}}k^2\,, \\
&\varepsilon_{\bk}^{\textrm{G}_0\textrm{W}_0,\textrm{e}}=\frac{\hbar^2 k^2}{2 m_{\textrm{e}}}+E^{\textrm{G}_0\textrm{W}_0}_{\textrm{Gap}}=\alpha_{\textrm{e}}k^2+E^{\textrm{G}_0\textrm{W}_0}_{\textrm{Gap}}
 \end{split}
\label{eq:disp}
\end{equation}
with the band gap $E^{\textrm{G}_0\textrm{W}_0}_{\textrm{Gap}}$ in G$_0$W$_0$ approximation. The band-gap renormalization due to the inner-shell electrons (TMDC and substrate) is given by the static SXCH self-energy terms in the first line of Eq.~(\ref{eq:BSE_HS}) and yields
\begin{equation}
 \begin{split}
 \Sigma_{\textrm{Gap}}^{\textrm{SXCH}}&=\Sigma_{\bk=\textbf{0}}^{\textrm{SXCH,e}}+\Sigma_{\bk=\textbf{0}}^{\textrm{SXCH,h}}\\
&=\frac{1}{\mathcal{A}}\sum_{\bk'}\big(W_{\bk'}^{\textrm{stat}}-W_{\bk'}^{\textrm{freest}}\big) \\
&=\frac{1}{\mathcal{A}}\sum_{\bk'}\frac{e^2}{2\varepsilon_0 |\bk'|}\left(\tilde{\varepsilon}^{-1}_{\bk'}-\frac{1}{1+r_0|\bk'|} \right) \\
&=\frac{1}{(2\pi)^2}\int_{0}^{2\pi}d\phi\int_{0}^{\infty}dk'\frac{e^2}{2\varepsilon_0 k'}\left(\tilde{\varepsilon}^{-1}_{k'}-\frac{1}{1+r_0 k'} \right) \\
&=-\frac{1}{2\pi}\frac{e^2}{2\varepsilon_0 r_0}\textrm{ln}\frac{1+\varepsilon_{\infty}}{2}\,,
 \end{split}
\label{eq:SXCH_gap}
\end{equation}
where we have introduced polar coordinates and performed the thermodynamic limit to calculate the momentum-space integral. 
To determine the dynamical band-gap renormalization, we have to evaluate
\begin{equation}
 \begin{split}
 &\Delta^{\textrm{dyn,e}}_{\bk=\textbf{0}}(E_{\textrm{Gap}})\\=&i\hbar\int_{-\infty}^{\infty}\frac{d\omega'}{2\pi}\frac{1}{\mathcal{A}}\sum_{\bk'}\frac{(1+n_{\textrm{B}}(\omega'))2i\textrm{Im}\,W^{\textrm{HS}}_{\bk'}(\omega')}{E_{\textrm{Gap}}-\varepsilon_{\textbf{0}}^{\textrm{h}}-\varepsilon_{\bk'}^{\textrm{e}}-\hbar\omega'+i\Gamma}\,,
 \end{split}
\label{eq:corr_diag_gap_keldysh}
\end{equation}
with quasi-particle energies $\varepsilon_{\bk}^{\textrm{e/h}}$ that are specified in the following, and the corresponding term for holes. The derivation of this expression involves no further approximation beyond the screened-ladder approximation to the interaction kernel of the BSE and the quasi-particle approximation to the single-particle Green functions. For the special case of $\hbar\omega=E_{\textrm{Gap}}$, the diagonal correlation $\Delta^{\textrm{dyn,e}}_{\bk=\textbf{0}}(\omega)$ is equivalent to the quasi-particle renormalization given by the single-particle self-energy $\Sigma^{\textrm{e}}_{\bk=\textbf{0}}(\varepsilon_{\textbf{0}}^{\textrm{e}})$ \cite{semkat_ionization_2009}. We expand the denominator in Eq.(\ref{eq:corr_diag_gap_keldysh}) as 
\begin{equation}
 \begin{split}
  E_{\textrm{Gap}}-\varepsilon_{\textbf{0}}^{\textrm{h}}&-\varepsilon_{\bk'}^{\textrm{e}} \\
 =E_{\textrm{Gap}}-(&\varepsilon_{\textbf{0}}^{\textrm{G}_0\textrm{W}_0,\textrm{h}}+\Sigma_{\textbf{0}}^{\textrm{SXCH,h}}+\Delta^{\textrm{dyn,h}}_{\textbf{0}}(E_{\textrm{Gap}})) \\
  -(&\varepsilon_{\bk'}^{\textrm{G}_0\textrm{W}_0,\textrm{e}}+\Sigma_{\bk'}^{\textrm{SXCH,e}}+\Delta^{\textrm{dyn,e}}_{\bk'}(E_{\textrm{Gap}})) \\
  \approx E_{\textrm{Gap}}-&E^{\textrm{G}_0\textrm{W}_0}_{\textrm{Gap}}-\alpha_{\textrm{e}}k'^2 \\
   -\Sigma_{\textbf{0}}^{\textrm{SXCH,h}}-&\Delta^{\textrm{dyn,h}}_{\textbf{0}}(E_{\textrm{Gap}})-\Sigma_{\textbf{0}}^{\textrm{SXCH,e}}-\Delta^{\textrm{dyn,e}}_{\textbf{0}}(E_{\textrm{Gap}})\\
   =-\alpha_{\textrm{e}}k'^2\,.
 \end{split}
\label{eq:energy_denom}
\end{equation}
In the third line we assumed that the static and dynamical renormalizations cause a rigid shift that has no momentum dependence and in the last line we made use of the self-consistency requirement that the total band gap should equal the band gap in G$_0$W$_0$ approximation plus all renormalizations. By introducing the explicit loss function (\ref{eq:im_total_invDF_delta}) and making use of the relation $n_{\textrm{B}}(-\omega)=-1-n_{\textrm{B}}(\omega)$ we arrive at 
\begin{equation}
 \begin{split}
 &\Delta^{\textrm{dyn,e}}_{k=0}(E_{\textrm{Gap}})=\frac{\hbar e^2}{4\pi\varepsilon_0}\frac{s^2}{2}\int_{0}^{\infty}dk'\frac{1}{\tilde{\varepsilon}_{k'}^2}\frac{1}{\omega_{1,k'}}
 \\ & \left\lbrace\frac{1+n_{\textrm{B}}(\omega_{1,k'})}{-\alpha_{\textrm{e}}k'^2-\hbar\omega_{1,k'}+i\Gamma}+\frac{n_{\textrm{B}}(\omega_{1,k'})}{-\alpha_{\textrm{e}}k'^2+\hbar\omega_{1,k'}+i\Gamma}\right\rbrace\,, \\
 &\Delta^{\textrm{dyn,h}}_{k=0}(E_{\textrm{Gap}})=\frac{\hbar e^2}{4\pi\varepsilon_0}\frac{s^2}{2}\int_{0}^{\infty}dk'\frac{1}{\tilde{\varepsilon}_{k'}^2}\frac{1}{\omega_{1,k'}}
 \\ & \left\lbrace\frac{1+n_{\textrm{B}}(\omega_{1,k'})}{-\alpha_{\textrm{h}}k'^2-\hbar\omega_{1,k'}+i\Gamma}+\frac{n_{\textrm{B}}(\omega_{1,k'})}{-\alpha_{\textrm{h}}k'^2+\hbar\omega_{1,k'}+i\Gamma}\right\rbrace\,.
 \end{split}
\label{eq:corr_diag_gap_final}
\end{equation}
In the case that the substrate dielectric function describes optical phonons, the real parts of these expressions corresponds to the polaron shift of the conduction and valence band, respectively \cite{haug_electron_1984}. From Eqs.(\ref{eq:corr_diag_gap_keldysh}) and (\ref{eq:energy_denom}) we also deduce that the renormalization of the quasi-particle band gap is not sensitive to screening at any specific energy scale. It rather averages over the loss function (inverse dielectric function) at all frequencies. However, a weighting factor approximately given by $(1+n_{\textrm{B}}(\omega))/\omega$ is involved, which clearly favors small frequencies. In this sense, low-frequency poles in the loss function are expected to affect the quasi-particle band gap more strongly than high-frequency poles. 
In a similar way as for the band gap, the diagonal and off-diagonal correlations can be evaluated at the exciton energy $E_{\textrm{X}}$ to calculate the renormalizations of the exciton:
\begin{equation}
 \begin{split}
 &\Delta^{\textrm{dyn,e}}_{k}(E_{\textrm{X}})\\=&\frac{\hbar e^2}{8\pi^2\varepsilon_0}\frac{s^2}{2}\int_{0}^{\infty}dk'\int_{0}^{2\pi}d\phi\frac{k'}{|k-k'|}\frac{1}{\tilde{\varepsilon}_{|k-k'|}^2}\frac{1}{\omega_{1,|k-k'|}}
 \\ & \bigg\lbrace\frac{1+n_{\textrm{B}}(\omega_{1,|k-k'|})}{-E_{\textrm{B}}-\alpha_{\textrm{h}}k^2-\alpha_{\textrm{e}}k'^2-\hbar\omega_{1,|k-k'|}+i\Gamma}\\&+\frac{n_{\textrm{B}}(\omega_{1,|k-k'|})}{-E_{\textrm{B}}-\alpha_{\textrm{h}}k^2-\alpha_{\textrm{e}}k'^2+\hbar\omega_{1,|k-k'|}+i\Gamma}\bigg\rbrace\,,  
 \\ 
 &\Xi^{\textrm{dyn,eh}}_{k,k'}(E_{\textrm{X}})\\=&\frac{\hbar e^2}{2\varepsilon_0}\frac{s^2}{2}\frac{1}{|k-k'|}\frac{1}{\tilde{\varepsilon}_{|k-k'|}^2}\frac{1}{\omega_{1,|k-k'|}}
 \\ & \bigg\lbrace\frac{1+n_{\textrm{B}}(\omega_{1,|k-k'|})}{-E_{\textrm{B}}-\alpha_{\textrm{h}}k^2-\alpha_{\textrm{e}}k'^2-\hbar\omega_{1,|k-k'|}+i\Gamma}\\&+\frac{n_{\textrm{B}}(\omega_{1,|k-k'|})}{-E_{\textrm{B}}-\alpha_{\textrm{h}}k^2-\alpha_{\textrm{e}}k'^2+\hbar\omega_{1,|k-k'|}+i\Gamma}\bigg\rbrace \\
 & + (\textrm{e}\leftrightarrow\textrm{h})\,,
 \end{split}
\label{eq:corr_x_final}
\end{equation}
where we introduced the exciton binding energy $E_{\textrm{B}}=E_{\textrm{Gap}}- E_{\textrm{X}}$. These renormalizations are due to scattering processes of excitons into unbound quasi-particles assisted by substrate plasmons. The scattering is particularly efficient when the energy transferred in these processes roughly corresponds to the exciton binding energy. Hence, unlike the band-gap renormalization, the exciton renormalization becomes resonant if the excitations in the loss function reach a certain energy scale, which is set by the exciton binding energy. 
We will study the consequences of this in the following section.
\par
By evaluating the correlation terms at the exciton energy, we removed the frequency dependence of the eigenvalue problem (\ref{eq:BSE_HS}), which can be written as 
\begin{equation}
 \begin{split}
 H(E_{\alpha})\ket{\psi_{\alpha}}=E_{\alpha}\ket{\psi_{\alpha}}
 \end{split}
\label{eq:eigen}
\end{equation}
with the effective Hamiltonian
\begin{equation}
 \begin{split}
 H_{kk'}(E_{\alpha})&= \big(\varepsilon_k^{\textrm{G}_0\textrm{W}_0,\textrm{e}}+\varepsilon_k^{\textrm{G}_0\textrm{W}_0,\textrm{h}}+\Sigma_k^{\textrm{SXCH,e}}+\Sigma_k^{\textrm{SXCH,h}}    
 \\&+\Delta^{\textrm{dyn,e}}_{k}(E_{\alpha})+\Delta^{\textrm{dyn,h}}_{k}(E_{\alpha})\big)\delta_{kk'} \\ 
 & -\frac{1}{\mathcal{A}} V_{k-k'}^{\textrm{stat}} -\frac{1}{\mathcal{A}} \Xi^{\textrm{dyn,eh}}_{k,k'}(E_{\alpha})  \,.
 \end{split}
\label{eq:eff_ham}
\end{equation}
However, $H(E_{\alpha})$ is non-Hermitian and thus has two sets of eigenstates. As shown in Ref.~\citenum{bornath_two-particle_1999}, this problem can be overcome by diagonalizing the Hermitian part of the Hamiltonian $\mathcal{H}H(E_{\alpha})$ under the condition that the off-diagonal matrix elements of the non-Hermitian part $\mathcal{A}H(E_{\alpha})$ are small compared to $\mathcal{H}H(E_{\alpha})$. To calculate bound exciton energies $E_{\textrm{X}}$ for a specific heterostructure, we numerically diagonalize the eigenvalue problem given by $\mathcal{H}H(E_{\textrm{X}})$ iteratively until self-consistency is reached. The exciton energy can then be compared to the quasi-particle band gap following directly from Eqs.(\ref{eq:SXCH_gap}) and (\ref{eq:corr_diag_gap_final}).
The matrix is set up in a basis of momentum states, where we can limit ourselves to the modulus of momenta due to the isotropy of the problem. On the off-diagonal, integration weights have to be included according to the thermodynamic limit $\frac{1}{\mathcal{A}}\sum_{\bk'}\rightarrow \frac{1}{(2\pi^2)}\int_{0}^{\infty}dk' k'\int_{0}^{2\pi}d\phi $. Moreover, singularities appear on the diagonal for $k=k'$, which 
we treat as described in \cite{haug_quantum_2004}.
\par
In the following section, we apply the theory developed above to monolayer MoS$_2$ placed on a three-dimensional substrate with varying carrier-doping density. 

\section{Results}

We describe the band structure of MoS$_2$ by equal electron and hole masses of $m_{\textrm{e}}=m_{\textrm{h}}=0.5\,m_0$ in agreement with Ref. \cite{berkelbach_theory_2013}. The polarizability of $\chi_{2\textrm{d}}=0.66$ nm (cgs units) is taken from the same reference. The parameters of the substrate are varied to study the influence of substrate plasmons on the TMDC electronic properties in a general way.
This may be realized by contacting a substrate electrically, as shown in Fig.~\ref{fig:scheme}, or by using indium-tin-oxide (ITO) crystals of different thickness that provide plasmon frequencies in the range of several hundred meV. \cite{chen_frequency-dependent_2010}
\begin{figure}[h!t]
\centering
\includegraphics[width=\columnwidth]{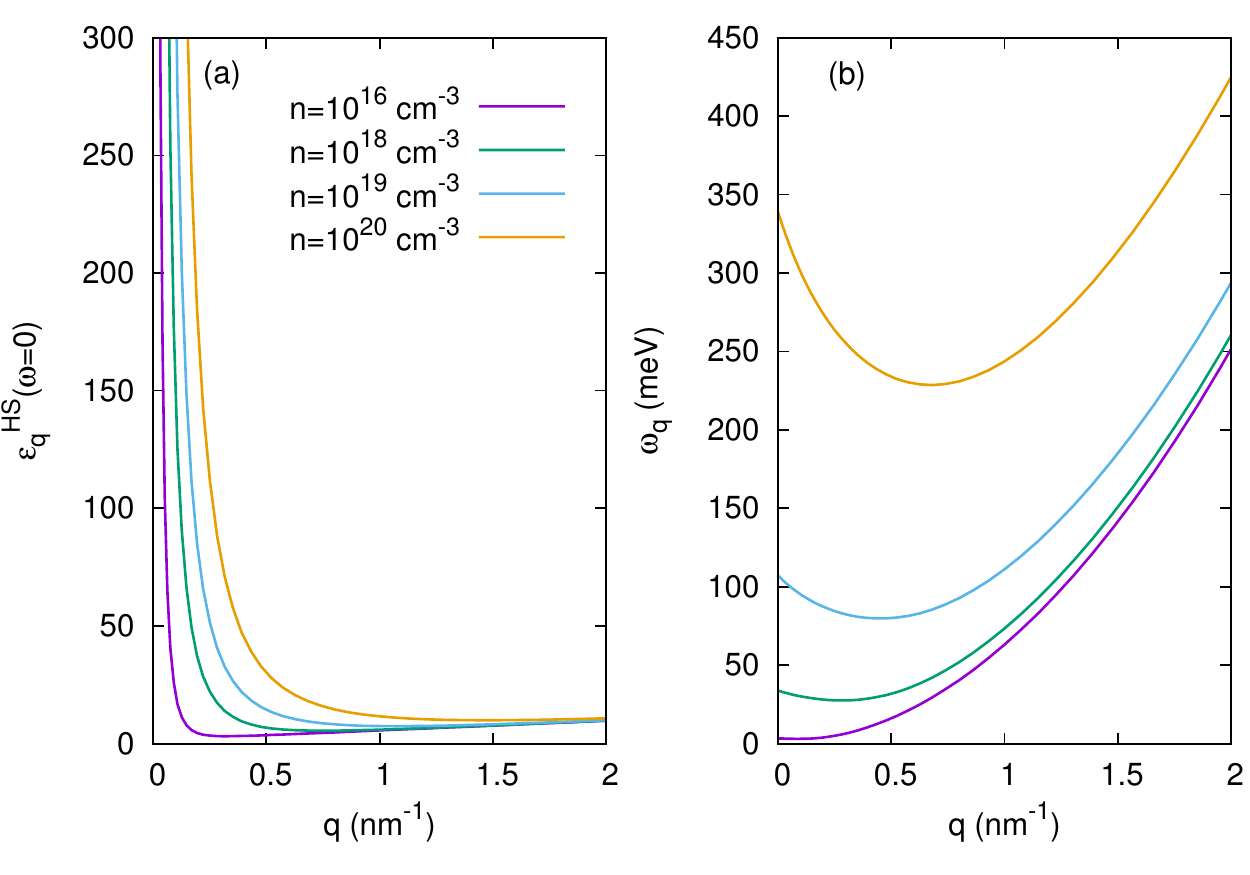}
\caption{\textbf{(a):} Heterostructure dielectric function in the static limit. \textbf{(b):} Resonance frequency $\omega_{1,\bq}$ of the loss function (imaginary part of the inverse heterostructure dielectric function). Both quantities are given for MoS$_2$ on a substrate with $\varepsilon_{\infty}=2$ and $m_{\textrm{s}}=0.4\,m_0$ at different carrier-doping densities $n$.}
%
\label{fig:resonance}
\end{figure}
\par

The total dielectric function $\varepsilon^{\textrm{HS}}_{\bq}(\omega)$ can be characterized by its static limit $\varepsilon^{\textrm{HS}}_{\bq}(\omega=0)$ and its dynamical part essentially defined by the momentum-dependent plasmon resonance at positive frequencies $\omega_{1,\bq}$. While the former exhibits a metallic screening behavior with a divergence at vanishing momentum (see Fig.~\ref{fig:resonance}(a)), the latter shows a parabolic behavior in momentum space. Both quantities are influenced by an increasing substrate doping level. The static part is simply enhanced with increasing carrier-doping density $n$. The plasmon resonance on the other hand develops a bowing around intermediate momenta for elevated $n$ due to the q-dependent electronic screening from the TMDC layer, see Eq.~(\ref{eq:resonance}). Together, these components encode all screening information of the heterostructure defined by the loss function $\textrm{Im}\,\varepsilon^{\textrm{HS},-1}_{\bq}(\omega)$ as given by Eq.~(\ref{eq:im_total_invDF_delta}).


\par
\begin{figure}[h!t]
\centering
\includegraphics[width=\columnwidth]{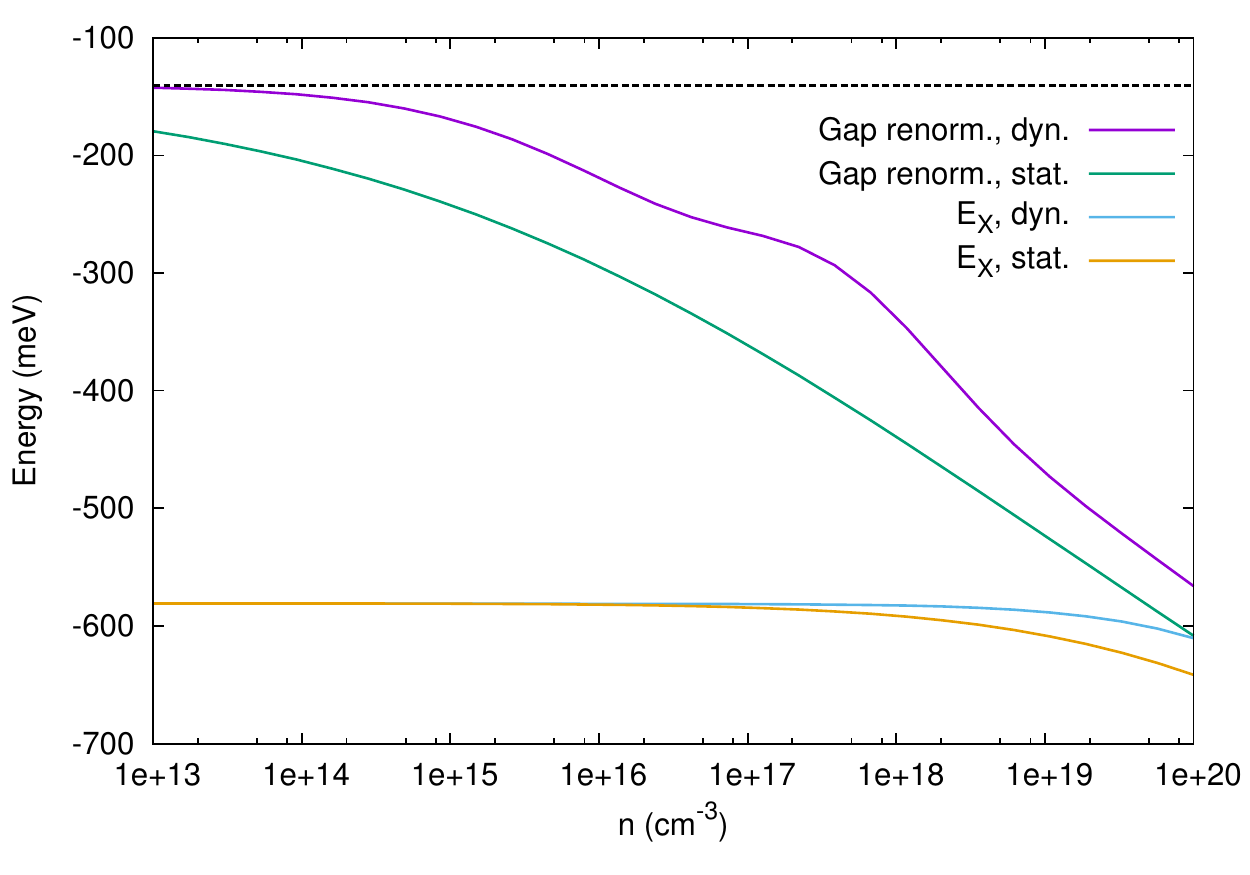}
\caption{Band-gap renormalization and 1s-exciton position relative to the unrenormalized band gap for MoS$_2$ on a substrate at $T=300$ K with $\varepsilon_{\infty}=2$, $m_{\textrm{s}}=0.4\,m_0$ and quasi-particle broadening $\Gamma=10$ meV, shown for different carrier-doping densities. The horizontal line shows the band-gap renormalization due to inner-shell electrons in substrate and TMDC layer. Results using the full dynamical dielectric function and a static approximation are compared.}
\label{fig:exciton_and_gap}
\end{figure}

By diagonalization of the dynamically screened BSE, we obtain the relative position of the lowest (1s) exciton of MoS$_2$ on a substrate with varying carrier density. The quasi-particle band-gap renormalization is given by the sum of the static and dynamical terms (\ref{eq:SXCH_gap}) and (\ref{eq:corr_diag_gap_final}).
For comparison, we also compute these quantities in static approximation 
as discussed in the Appendix, 
thereby neglecting all dynamical correlations. The results are shown in Fig.~\ref{fig:exciton_and_gap}.

The SXCH-renormalizations due to inner electrons in the substrate and TMDC layer given by Eq.~(\ref{eq:SXCH_gap}) yield a density-independent gap shift of about $140$ meV. In both theories, we find an increasing gap shift with increasing substrate doping. However, the static calculation overestimates the gap renormalization and also neglects a nontrivial S-shape-like dependence on the carrier density which is found in the dynamical calculation. The latter is a result of the correlation defined in Eq.~(\ref{eq:corr_diag_gap_final}) being dependent on the loss-function resonance $\omega_{1,\bq}$, which is neglected in the static calculation.
The position of the 1s-exciton relative to the unrenormalized band gap is stable over a wide range of carrier densities. It shows a redshift at elevated densities that is less pronounced and sets in significantly later in the case of dynamical screening.  
The compensation of different renormalization effects leading to a relatively stable exciton position in TMDCs is known from previous studies based on statically screened interaction \cite{berkelbach_theory_2013, steinhoff_influence_2014}.
It has also been discussed for the case of dynamical screening of excitons \cite{gao_dynamical_2016}. 
The compensation is better in the case of a dynamically screened interaction. However, due to the frequency-dependence of correlations, the picture of excitons being ``pinned'' to the renormalized quasi-particle band gap and shifting with it breaks down in general. In the dynamical calculation, the exciton position is determined by the competition of diagonal and off-diagonal correlation terms evaluated at the exciton energy, while the band-gap shift is sensitive to the diagonal dephasing at the band-gap energy. Only for a static interaction, both quantities are directly connected.
As a net result of the stable exciton position, the exciton binding energy $E_{\textrm{B}}=E_{\textrm{Gap}}- E_{\textrm{X}}$ is tuned by substrate doping over several hundred meV.
The exciton Mott transition, where the bound exciton disappears as its binding energy is reduced to zero by many-body renormalizations, takes place above a substrate carier density $n=10^{20}$ cm$^{-3}$. Note that this critical density is very low compared to typical densities of free carriers on the order of $10^{23}$ cm$^{-3}$ found in metals like gold or aluminum. 

\par
\begin{figure}[h!t]
\centering
\includegraphics[width=\columnwidth]{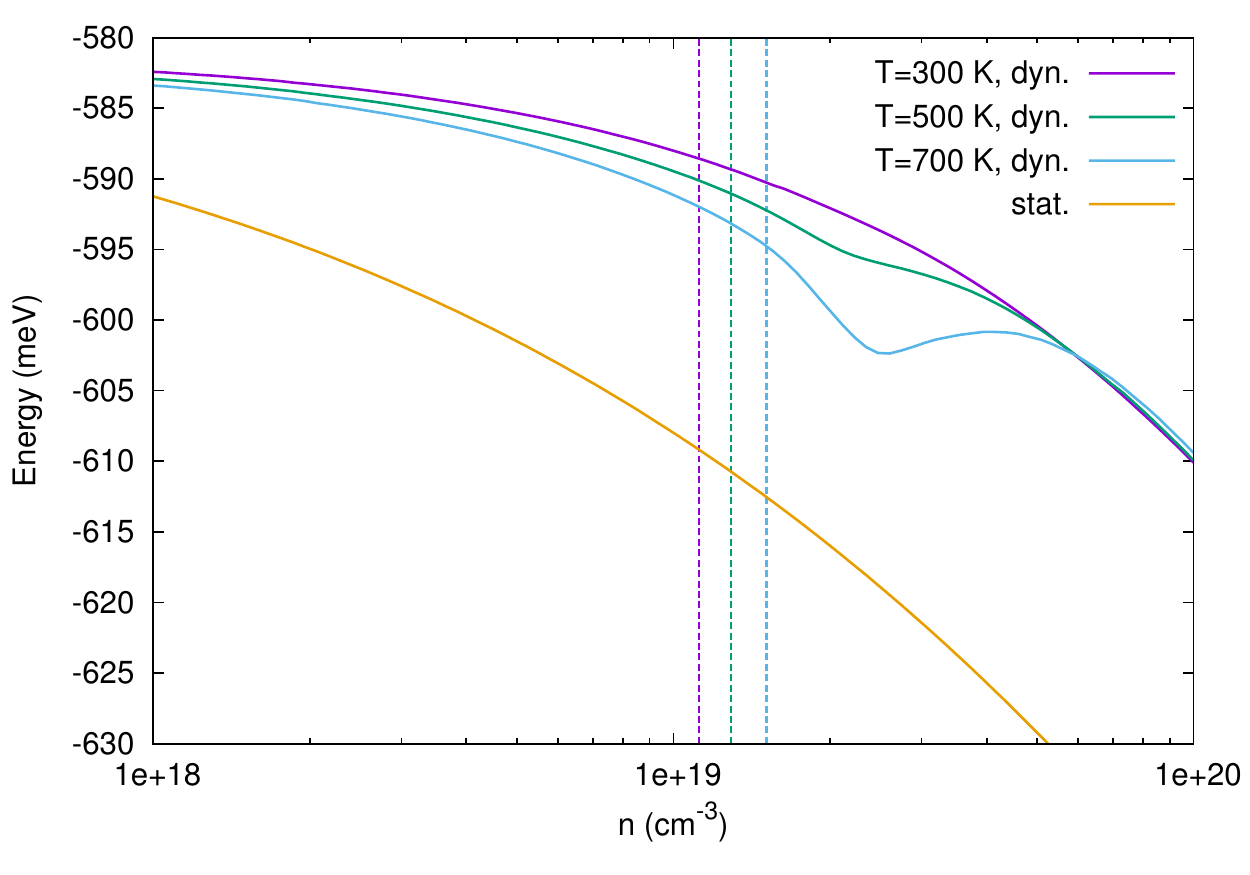}
\caption{1s-exciton position relative to the unrenormalized band gap for MoS$_2$ on a substrate at different temperatures with $\varepsilon_{\infty}=2$, $m_{\textrm{s}}=0.4\,m_0$ and quasi-particle broadening $\Gamma=10$ meV, shown for different carrier-doping densities. The vertical dashed lines show the density where the loss function resonance at zero momentum becomes resonant with the exciton binding energy for all three temperatures. The resonance shifts to larger densities with increasing temperature. For comparison, the result obtained from the static approximation to the dielectric function is shown.}
\label{fig:resonance_renorm}
\end{figure}
As we have discussed in the previous section, the correlation terms at the exciton energy (\ref{eq:corr_x_final}) in principle exhibit a resonance when the excitations in the loss function become comparable to the exciton binding energy. As Fig.~\ref{fig:exciton_and_gap} shows, no resonance is observed in the exciton position though, which is due to two reasons. The resonance appears in the second term of the correlations, which are proportional to the population of plasmons $n_{\textrm{B}}(\omega_{1,k-k'})$. This term corresponds to scattering of one of the carriers in the exciton between its actual state $\ket{k}$ and a state $\ket{k'}$ under the absorption of a plasmon with frequency $\omega_{1,k-k'}$, while the other carrier plays the role of a spectator. At room temperature, the population of plasmons at energies of several hundred meV is low, so that this scattering becomes very inefficient and hence the renormalization it causes is small even in the case of resonance.
The same resonance would also be driven by an emission event proportional to $1+n_{\textrm{B}}(\omega_{1,k-k'})$, which requires occupancy of the (electron or hole) state $\ket{k'}$. Hence, carrier doping of the TMDC layer itself might be beneficial to develop a resonance. At the same time, this carrier doping introduces additional many-body renormalizations that are beyond the scope of this paper.
The resonance appears in both, the diagonal and off-diagonal correlation simultaneously, leading to efficient compensation of any renormalization. Nevertheless, we demonstrate a resonance effect by increasing the lattice temperature up to $700$ K as shown in Fig.~\ref{fig:resonance_renorm}.
We find a resonance in the exciton position emerging with increasing temperature beyond $T=300$ K. As expected, the resonance roughly correlates with the excitation energy in the loss function at zero momentum, $\omega_{1,0}$, being equal to the exciton binding energy. Due to the momentum-dependence of $\omega_{1,q}$, this is, however, just an estimate for the density range of interest. For comparison, we also give the result obtained in static approximation to the heterostructure dielectric function. Due to the lacking frequency dependence, no resonance shows up and the exciton position does not depend on temperature. 
\par
To give a more general discussion of the observed effects, we also study the dependence of band-gap and exciton renormalizations on the model parameters, which are the high-frequency dielectric constant due to inner electrons in the substrate $\varepsilon_{\infty}$, the effective mass of substrate carriers $m_{\textrm{s}}$ and the quasi-particle broadening $\Gamma$. The data is shown and discussed in the Appendix.

\section{Conclusion}

We have derived a BSE for excitons in atomically thin TMDC semiconductors on a three-dimensional substrate including effects of frequency-dependent screening due to bosonic excitations in the substrate. The theory is applied to monolayer MoS$_2$ on substrates with varying carrier-doping density and corresponding plasmonic properties. We find the 1s-exciton to be spectrally stable over a wide doping range before showing a red shift of up to several ten meV. The red shift is systematically smaller than estimated by a theory based on static screening. While the band-gap renormalization shows a nontrivial S-shape behavior depending on the density, an exciton-plasmon resonance for plasmons that match the exciton binding energy is only observed at high temperatures. This is due to the low thermal-equilibrium population of high-energy plasmons and efficient compensation between different renormalization effects. Unlike suggested by the often applied picture of statically screened Coulomb interaction, the energies of excitons and quasi-particle band gap are decoupled, as they are sensitive to correlations at different frequencies. The quasi-particle band gap is not sensitive to screening at any specific frequency range. Instead we find that the band-gap renormalization averages over the inverse dielectric function at all frequencies with a weight factor approximately given by $1/\omega$ at low temperatures that clearly favors low plasmon frequencies. Hence resonances in the loss function at frequencies on the order of the quasi-particle band gap yield no essential contribution to exciton and band-gap renormalizations.
\par
Our approach can be extended to arbitrary heterostructures by choosing an appropriate macroscopic model as discussed in detail in Refs.\citenum{rosner_wannier_2015,latini_excitons_2015,meckbach_influence_2018,florian_dielectric_2018}. Moreover, two-dimensional substrates, such as additional layers of TMDCs, can be studied. The two-dimensional plasmons in these materials behave very differently from the three-dimensional plasmon studied in this paper \cite{groenewald_valley_2016} and may open additional possibilities to tailor quasi-particle band gaps and excitons. Beyond that, our results are a first step towards a microscopic understanding of how TMDC excitons can be manipulated by means of more complex plasmonic nanostructures.


$\vspace{5mm}$

\textbf{Acknowledgement}

M.R. would like to thank the Alexander von Humboldt Foundation for support.

\section{Appendix}

\subsection{Derivation of the BSE for heterostructures}

We first give the BSE in general form before we derive a particular version that can be applied to the heterostructure shown in Fig.~\ref{fig:scheme}. In the electron-hole (e-h) picture, assuming one electron and hole band, respectively, the screened-ladder BSE has the form \cite{bornath_two-particle_1999, kremp_quantum_2005}
\begin{equation}
 \begin{split}
  & (\varepsilon_{\bk}^{\textrm{0,e}}+\varepsilon_{\bk}^{\textrm{0,h}}+\Delta^{\textrm{e}}_{\bk}(\omega)+\Delta^{\textrm{h}}_{\bk}(\omega)-\hbar\omega)\psi^{\textrm{eh}}_{\bk}(\omega) \\
  &-(1-f_{\bk}^{\textrm{e}}-f_{\bk}^{\textrm{h}})\frac{1}{\mathcal{A}}\sum_{\bk'}V^{\textrm{eh}}_{\bk\bk'}\psi^{\textrm{eh}}_{\bk'}(\omega) \\ 
  &-\frac{1}{\mathcal{A}}\sum_{\bk'}\Xi^{\textrm{eh}}_{\bk\bk'}(\omega)\psi^{\textrm{eh}}_{\bk'}(\omega)=0\,.
 \end{split}
\label{eq:BSE_general}
\end{equation}
Here, $\psi^{\textrm{eh}}_{\bk}(\omega)$ denotes the two-particle spectral function, $\varepsilon_{\bk}^{\textrm{0,e/h}}$ are electron and hole energies containing renormalizations at most on a mean-field level, 
\begin{equation}
 \begin{split}
&V^{\textrm{eh}}_{\bk\bk'}=\\ &\int\int\textbf{dr}^3\textbf{dr}'^3(\Phi_{\bk}^{\textrm{e}}(\textbf{r}))^*(\Phi^{\textrm{h}}_{\bk'}(\textbf{r}'))^*\frac{1}{|\textbf{r}-\textbf{r}'|}\Phi^{\textrm{h}}_{\bk}(\textbf{r}')\Phi^{\textrm{e}}_{\bk'}(\textbf{r})
 \end{split}
\label{eq:bare_Coul_app}
\end{equation}
is the bare electron-hole interaction and $\mathcal{A}$ is the area of the TMDC layer. We assume in the following that no electron-hole pairs are excited, setting the electron and hole occupancies $f_{\textbf{k}}^{\textrm{e/h}}=0$.
All correlation effects are contained in the expressions
\begin{equation}
 \begin{split}
 \Delta^{\textrm{e}}_{\bk}(\omega)&=i\hbar\int_{-\infty}^{\infty}\frac{d\omega'}{2\pi}\frac{1}{\mathcal{A}}\sum_{\bk'}\frac{(1+n_{\textrm{B}}(\omega'))2i\textrm{Im}\,W^{\textrm{ee}}_{\bk\bk'}(\omega')}{\hbar\omega-\varepsilon_{\bk}^{\textrm{h}}-\varepsilon_{\bk'}^{\textrm{e}}-\hbar\omega'+i\Gamma}\,, \\
 \Delta^{\textrm{h}}_{\bk}(\omega)&=i\hbar\int_{-\infty}^{\infty}\frac{d\omega'}{2\pi}\frac{1}{\mathcal{A}}\sum_{\bk'}\frac{(1+n_{\textrm{B}}(\omega'))2i\textrm{Im}\,W^{\textrm{hh}}_{\bk\bk'}(\omega')}{\hbar\omega-\varepsilon_{\bk}^{\textrm{e}}-\varepsilon_{\bk'}^{\textrm{h}}-\hbar\omega'+i\Gamma}
 \end{split}
\label{eq:corr_diag}
\end{equation}
and
\begin{equation}
 \begin{split}
 \Xi^{\textrm{eh}}_{\bk\bk'}(\omega)&=i\hbar\int_{-\infty}^{\infty}\frac{d\omega'}{2\pi}\frac{(1+n_{\textrm{B}}(\omega'))2i\textrm{Im}\,W^{\textrm{eh}}_{\bk\bk'}(\omega')}{\hbar\omega-\varepsilon_{\bk}^{\textrm{h}}-\varepsilon_{\bk'}^{\textrm{e}}-\hbar\omega'+i\Gamma} \\ &+(\textrm{e}\leftrightarrow \textrm{h})\,,
 \end{split}
\label{eq:corr_offdiag}
\end{equation}
which we term diagonal and off-diagonal correlations in the following. The correlations are essentially of the GW form, as the screened-ladder approximation for the two-particle Green function is consistent with the GW approximation on the single-particle level. \cite{baym_conservation_1961, kremp_quantum_2005}
Hence the quasi-particle energies $\varepsilon_{\bk}^{\textrm{e/h}}$ are usually calculated on a GW-level. To derive the expressions (\ref{eq:corr_diag}) and (\ref{eq:corr_offdiag}), thermal equilibrium relations for the propagators of the screened Coulomb interaction are used \cite{kremp_quantum_2005}:
\begin{equation}
\begin{split} 
  & W^{<,\textrm{ab}}_{\bk\bk'}(\omega)=n_{\textrm{B}}(\omega)2i\textrm{Im}\,W^{\textrm{ab,ret}}_{\bk\bk'}(\omega), \\
  & W^{>,\textrm{ab}}_{\bk\bk'}(\omega)=(1+n_{\textrm{B}}(\omega))2i\textrm{Im}\,W^{\textrm{ab,ret}}_{\bk\bk'}(\omega)\,.
    \label{eq:thermal_plasmon_prop}
\end{split}
\end{equation}
$n_{\textrm{B}}(\omega)$ is the Bose distribution function describing the thermal-equilibrium population of bosonic excitations. In the following we do not label retarded quantities explicitely as there is no confusion possible. $W^{\textrm{ab}}_{\bk\bk'}(\omega)$ is the retarded screened interaction between carriers in bands a and b given by
\begin{equation}
 \begin{split}
 W^{\textrm{ab}}_{\bk\bk'}(\omega)=V^{\textrm{ab}}_{\bk\bk'}\varepsilon^{-1}_{\bk-\bk'}(\omega)
 \end{split}
\label{eq:carrier_carrier_app}
\end{equation}
with a general macroscopic retarded dielectric function $\varepsilon^{-1}_{\bq}(\omega)$. $\Gamma$ is a phenomenological quasi-particle broadening.
\par
We proceed by specializing the BSE for the heterostructure of TMDC layer and substrate shown in Fig.~\ref{fig:scheme}. This means in particular that we replace the general screened interaction $W$ by an interaction $W^{\textrm{HS}}$ between carriers in the TMDC in the presence of the dielectric screening of the full heterostructure. First-principle electronic structure calculations are usually performed for freestanding TMDC layers, while the material-realistic treatment of bulk substrates is challenging. Hence, we would like to quantify the additional effects caused by substrate screening relative to the electronic properties, such as band gap and exciton position, of the freestanding TMDC.
To this end, we split the screened carrier-carrier interaction into a part which is purely due to the freestanding TMDC layer and a part that contains the rest. At the same time we assume that the single-particle Green function that is used to evaluate the self-energy is not modified in the presence of the substrate. Schematically, the GW self-energy of the heterostructure is written as
\begin{equation}
 \begin{split}
 \Sigma^{\textrm{HS,GW}}&=GW^{\textrm{HS}} \\ &=GW^{\textrm{freest}}+G(W^{\textrm{HS}}-W^{\textrm{freest}})=G\Delta W\,,
 \end{split}
\label{eq:GdW}
\end{equation}
following the GdW approach \cite{rohlfing_electronic_2010, winther_band_2017}. In the spirit of a DFT+G$_0$W$_0$ calculation, we assume that the diagonal correlation term (\ref{eq:corr_diag}) is evaluated using the carrier-carrier interaction of the freestanding layer, the energies $\varepsilon_{\bk}^{\textrm{0,e/h}}$ and a quasi-particle ansatz. This yields G$_0$W$_0$-corrections to the energies $\varepsilon_{\bk}^{\textrm{0,e/h}}$ in Eq.~(\ref{eq:BSE_general}):
\begin{equation}
 \begin{split}
 \varepsilon_{\bk}^{\textrm{G}_0\textrm{W}_0,\textrm{e/h}}=\varepsilon_{\bk}^{\textrm{0,e/h}}+\Delta_{\bk}^{\textrm{G}_0\textrm{W}_0,\textrm{e/h}}\,.
 \end{split}
\label{eq:GdW_energies}
\end{equation}
We evaluate the off-diagonal correlation term (\ref{eq:corr_offdiag}) including $W^{\textrm{freest}}$ assuming that the frequency-dependence of carrier-carrier interaction inside the TMDC layer is weak at low energies. Hence we can take the static limit of this term as shown in the following section, which yields
\begin{equation}
 \begin{split}
 \Xi^{\textrm{freest,eh}}_{\bk\bk'}(\omega)\approx W_{\bk\bk'}^{\textrm{freest,eh}}-V^{\textrm{eh}}_{\bk\bk'}\,
 \end{split}
\label{eq:corr_offdiag_static}
\end{equation}
with $W_{\bk\bk'}^{\textrm{freest,eh}}=V^{\textrm{eh}}_{\bk\bk'}\varepsilon^{\textrm{freest},-1}_{\bk-\bk'}(\omega=0)$.
\par
The next step is to evaluate the correlation terms using the difference part of the carrier-carrier interaction from Eq.~(\ref{eq:GdW}), $\Delta W=W^{\textrm{HS}}-W^{\textrm{freest}}$. While we may again assume that the freestanding-layer part of interaction can be treated in the static limit, we have to be careful with the interaction that involves the full heterostructure. As we have derived in the section about dielectric functions, a suitable model for the interaction of TMDC carriers with excitations in the substrate is the linearized Keldysh potential given by Eq.~(\ref{eq:carrier_carrier}) with the inverse dielectric function given by Eq.~(\ref{eq:total_invDF}). The second term of the inverse dielectric function stemming from electron-boson interaction is clearly frequency-dependent and can be inserted into the correlation integrals in a straightforward fashion. The first term, however, has no proper frequency dependence and would give no contribution to the loss function that enters the correlation terms. We therefore assume that it represents the static limit 
\begin{equation}
 \begin{split}
  \varepsilon^{\textrm{HS,stat},-1}_{\bq}=\tilde{\varepsilon}^{-1}_{\bq}
 \end{split}
\label{eq:V_stat}
\end{equation}
of some proper inverse dielectric function, for which we can follow the same procedure as for the freestanding-layer interaction.
The static limit of the correlations involving $W^{\textrm{freest}}$ and $W^{\textrm{stat}}=V \varepsilon^{\textrm{HS,stat},-1}$ is performed taking into account the filled valence and empty conduction bands of the TMDC described by the occupancies $f_{\bk}^{\textrm{v}}=1$ and $f_{\bk}^{\textrm{c}}=0$. For the diagonal correlation we obtain, neglecting inter-band exchange terms,
\begin{equation}
 \begin{split}
 \Delta^{\textrm{freest,c}}_{\bk}(\omega)& \approx \frac{1}{\mathcal{A}}\sum_{\bk'}V^{\textrm{cc}}_{\bk\bk'}f_{\bk'}^{\textrm{c}}-\frac{1}{\mathcal{A}}\sum_{\bk'}W_{\bk\bk'}^{\textrm{freest,cc}}f_{\bk'}^{\textrm{c}} \\
 &+\frac{1}{2\mathcal{A}}\sum_{\bk'}\big(W_{\bk\bk'}^{\textrm{freest,cc}}-V^{\textrm{cc}}_{\bk\bk'}\big)\,, \\
 \Delta^{\textrm{stat,c}}_{\bk}(\omega)& \approx \frac{1}{\mathcal{A}}\sum_{\bk'}V^{\textrm{cc}}_{\bk\bk'}f_{\bk'}^{\textrm{c}}-\frac{1}{\mathcal{A}}\sum_{\bk'}W_{\bk\bk'}^{\textrm{stat,cc}}f_{\bk'}^{\textrm{c}} \\
 &+\frac{1}{2\mathcal{A}}\sum_{\bk'}\big(W_{\bk\bk'}^{\textrm{stat,cc}}-V^{\textrm{cc}}_{\bk\bk'}\big) 
 \end{split}
\label{eq:corr_diag_static}
\end{equation}
while the off-diagonal correlation yields
\begin{equation}
 \begin{split}
 \Xi^{\textrm{stat,eh}}_{\bk\bk'}(\omega)\approx W_{\bk\bk'}^{\textrm{stat,eh}}-V^{\textrm{eh}}_{\bk\bk'}\,
 \end{split}
\label{eq:corr_offdiag_static_2}
\end{equation}
with $\Xi^{\textrm{freest}}$ already given in Eq.~(\ref{eq:corr_offdiag_static}). We can now plug the G$_0$W$_0$-corrections (\ref{eq:GdW_energies}) as well as the static limits of correlations given by (\ref{eq:corr_offdiag_static}), (\ref{eq:corr_diag_static}) and (\ref{eq:corr_offdiag_static_2}) into the general BSE (\ref{eq:BSE_general}) to obtain the BSE for the heterostructure:
\begin{equation}
 \begin{split}
  & (\varepsilon_{\bk}^{\textrm{G}_0\textrm{W}_0,\textrm{e}}+\varepsilon_{\bk}^{\textrm{G}_0\textrm{W}_0,\textrm{h}}+\Sigma_{\bk}^{\textrm{SXCH,e}}+\Sigma_{\bk}^{\textrm{SXCH,h}}    \\&+\Delta^{\textrm{dyn,e}}_{\bk}(\omega)+\Delta^{\textrm{dyn,h}}_{\bk}(\omega)-\hbar\omega)\psi^{\textrm{eh}}_{\bk}(\omega) \\
  &-\frac{1}{\mathcal{A}}\sum_{\bk'}W_{\bk\bk'}^{\textrm{stat,eh}}\psi^{\textrm{eh}}_{\bk'}(\omega)\\&-\frac{1}{\mathcal{A}}\sum_{\bk'}\Xi^{\textrm{dyn,eh}}_{\bk\bk'}(\omega)\psi^{\textrm{eh}}_{\bk'}(\omega)=0\,.
 \end{split}
\label{eq:BSE_HS_appendix}
\end{equation}
The diagonal and off-diagonal correlations now contain only contributions due to the dynamical or frequency-dependent part of the inverse total dielectric function of the full heterostructure:
\begin{equation}
 \begin{split}
W^{\textrm{HS,ab}}_{\bk\bk'}(\omega)=V^{\textrm{ab}}_{\bk\bk'}\varepsilon^{\textrm{HS},-1}_{\bk-\bk'}(\omega)\,.
 \end{split}
\label{eq:W_HS}
\end{equation}
A statically screened electron-hole interaction term is defined by
\begin{equation}
 \begin{split}
W_{\bk\bk'}^{\textrm{stat,eh}}=V^{\textrm{eh}}_{\bk\bk'}\varepsilon^{\textrm{HS,stat},-1}_{\bk-\bk'}=V^{\textrm{eh}}_{\bk\bk'}\tilde{\varepsilon}^{-1}_{\bk-\bk'}
 \end{split}
\label{eq:static_e_h_term_appendix}
\end{equation}
and screened-exchange (SX) as well as Coulomb-hole (CH) terms \cite{haug_quantum_2004, steinhoff_influence_2014} describing static band-structure renormalizations are given by
\begin{equation}
 \begin{split}
\Sigma_{\bk}^{\textrm{SXCH,e}}&=\Sigma_{\bk}^{\textrm{SXCH,c}} \\ &=\frac{1}{2\mathcal{A}}\sum_{\bk'}\big(W_{\bk\bk'}^{\textrm{stat,ee}}-W_{\bk\bk'}^{\textrm{freest,ee}}\big) \\
\Sigma_{\bk}^{\textrm{SXCH,h}}&=-\Sigma_{\bk}^{\textrm{SXCH,v}}\\ &=\frac{1}{2\mathcal{A}}\sum_{\bk'}\big(W_{\bk\bk'}^{\textrm{stat,hh}}-W_{\bk\bk'}^{\textrm{freest,hh}}\big) \,.
 \end{split}
\label{eq:SXCH}
\end{equation}

\subsection{Static limit of BSE}

In this section, we discuss the static limit of the correlation terms entering the dynamically screened BSE (\ref{eq:BSE_HS}). This limit can be systematically derived by assuming that any excitation energy into pairs of free (quasi-)particles, $\hbar\omega-\varepsilon^{\textrm{a}}_{\bk}-\varepsilon^{\textrm{b}}_{\bk'}$, in the correlation integrals (\ref{eq:corr_diag}) and (\ref{eq:corr_offdiag}) is small compared to characteristic energies $\hbar\omega'$ occurring in the dielectric function. \cite{bornath_two-particle_1999, kremp_quantum_2005} Then we obtain for example
\begin{equation}
\begin{split}
\Delta^{\textrm{e}}_{\bk}(\omega)&\approx -i\hbar\int_{-\infty}^{\infty}\frac{d\omega'}{2\pi} \frac{1}{\mathcal{A}}\sum_{\bk'} \\
    &\frac{(1-f^{\textrm{e}}_{\bk'}+n_{\textrm{B}}(\omega'))2i V^{\textrm{ee}}_{\bk\bk'}\textrm{Im}\,\varepsilon^{-1}_{\bk-\bk'}(\omega')}{\hbar\omega'}
    \,.
\label{eq:Delta_approx1}
\end{split}
\end{equation}
We have written the diagonal correlation in the electron-hole picture, which is suitable to describe a semiconductor in its ground state as devoid of any particles, including a finite carrier population $f^{\textrm{e}}_{\bk}$. We may however transform the correlation to the picture of valence- and conduction-band electrons by using 
\begin{equation}
\begin{split}
f^{\textrm{e}}_{\bk}&=f^{\textrm{c}}_{\bk}, \\
f^{\textrm{h}}_{\bk}&=1-f^{\textrm{v}}_{\bk}, \\
\varepsilon^{\textrm{e}}_{\bk}&=\varepsilon^{\textrm{c}}_{\bk}, \\
\varepsilon^{\textrm{h}}_{\bk}&=-\varepsilon^{\textrm{v}}_{\bk}
    \,.
\label{eq:e_h_picture}
\end{split}
\end{equation}
In this case, care has to be taken to avoid double-counting of many-body interactions that are already included in the ground-state band structure.
We use the relation 
\begin{equation}
\begin{split}
\mathcal{P} \int_{-\infty}^{\infty}\frac{d\omega'}{\pi}\frac{\textrm{Im}\,\varepsilon^{-1}_{\bq}(\omega')}{\omega'-\omega}=\textrm{Re}\,\varepsilon^{-1}_{\bq}(\omega)-1
\label{eq:KK}
\end{split}
\end{equation}
with the Cauchy principal value $\mathcal{P}$, corresponding to the dispersion relation for the electronic susceptibility, for $\omega=0$. Since $\textrm{Im}\,\varepsilon^{-1}_{\bq}(\omega)$ is an odd function of $\omega$, the integrand has no pole at $\omega'=0$ and the principal value becomes a regular integral. Furthermore, we use the fact that $n_{\textrm{B}}(\omega)+\frac{1}{2}$ is an odd function of $\omega$ as well, hence
\begin{equation}
\begin{split}
\int_{-\infty}^{\infty}\frac{d\omega'}{\pi}\frac{\textrm{Im}\,\varepsilon^{-1}_{\bq}(\omega')(n_{\textrm{B}}(\omega')+\frac{1}{2})}{\omega'}=0\,.
\label{eq:integral2}
\end{split}
\end{equation}
Combining Eq.~(\ref{eq:Delta_approx1}) with (\ref{eq:KK}) and (\ref{eq:integral2}), the diagonal correlation becomes
\begin{equation}
\begin{split}
\Delta^{\textrm{e}}_{\bk}(\omega)&\approx \frac{1}{\mathcal{A}}\sum_{\bk'}\big( \frac{1}{2}-f_{\bk'}^{\textrm{e}}\big)\left[W^{\textrm{ee}}_{\bk\bk'}(\omega=0)- V^{\textrm{ee}}_{\bk\bk'}\right]
\label{eq:Delta_approx2}
\end{split}
\end{equation}
with $W^{\textrm{ee}}_{\bk\bk'}(\omega=0)=V^{\textrm{ee}}_{\bk\bk'} \varepsilon^{-1}_{\bk-\bk'}(\omega=0) $. Similarly, the off-diagonal correlation is given by
\begin{equation}
\begin{split}
\Xi_{\bk\bk'}^{\textrm{eh}}(\omega)&\approx \big(1-f_{\bk}^{\textrm{h}}-f_{\bk}^{\textrm{e}}\big)\left[W^{\textrm{eh}}_{\bk\bk'}(\omega=0)- V^{\textrm{eh}}_{\bk\bk'}\right]\,.
\label{eq:Veff_approx2}
\end{split}
\end{equation}

\subsection{Parameter dependence of band-gap and exciton renormalization}
\begin{figure}[h!t]
\centering
\includegraphics[width=\columnwidth]{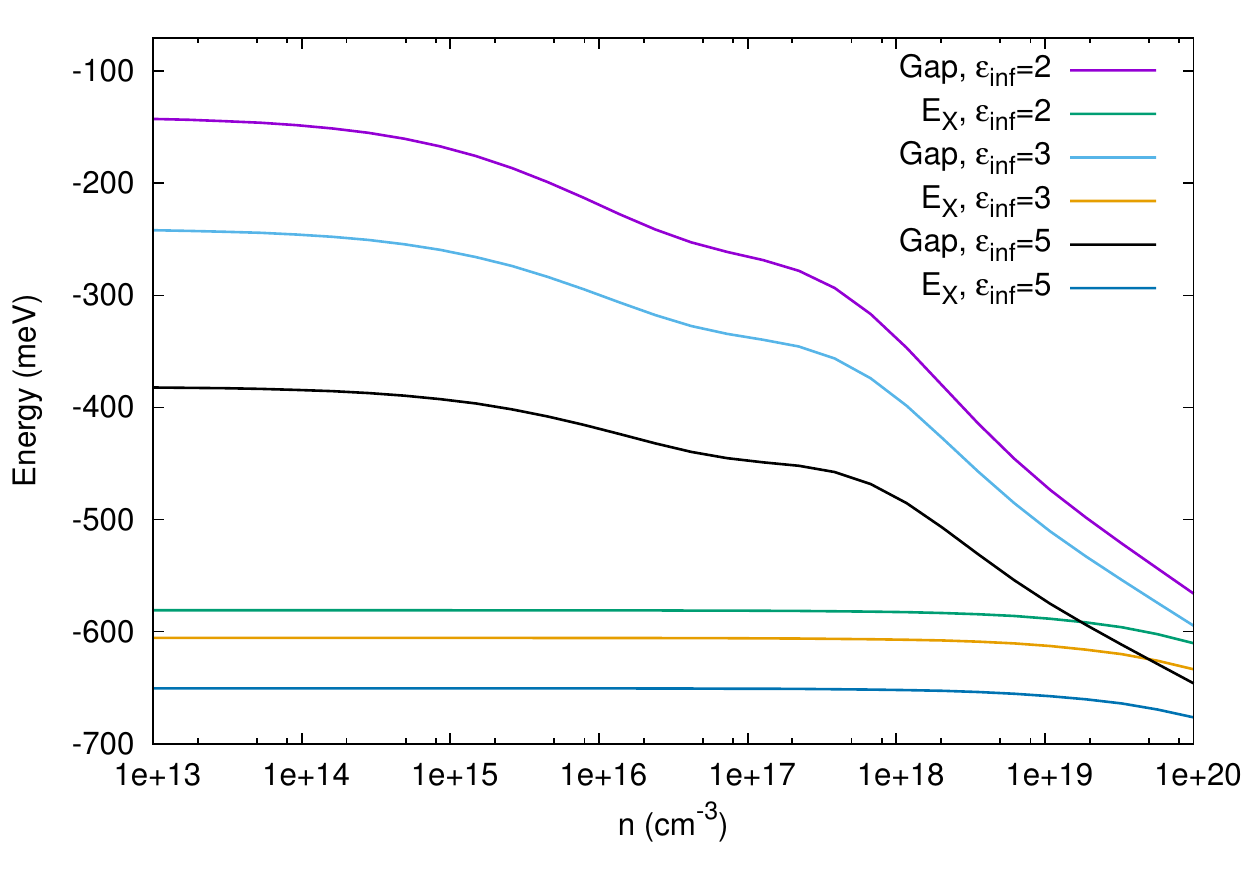}
\caption{Band-gap renormalization and 1s-exciton position relative to the unrenormalized band gap for MoS$_2$ on a substrate at $T=300$ K with $m_{\textrm{s}}=0.4\,m_0$ and quasi-particle broadening $\Gamma=10$ meV, shown for different carrier-doping densities and dielectric constants $\varepsilon_{\infty}$.}
\label{fig:eps_inf_dep}
\end{figure}
\begin{figure}[h!t]
\centering
\includegraphics[width=\columnwidth]{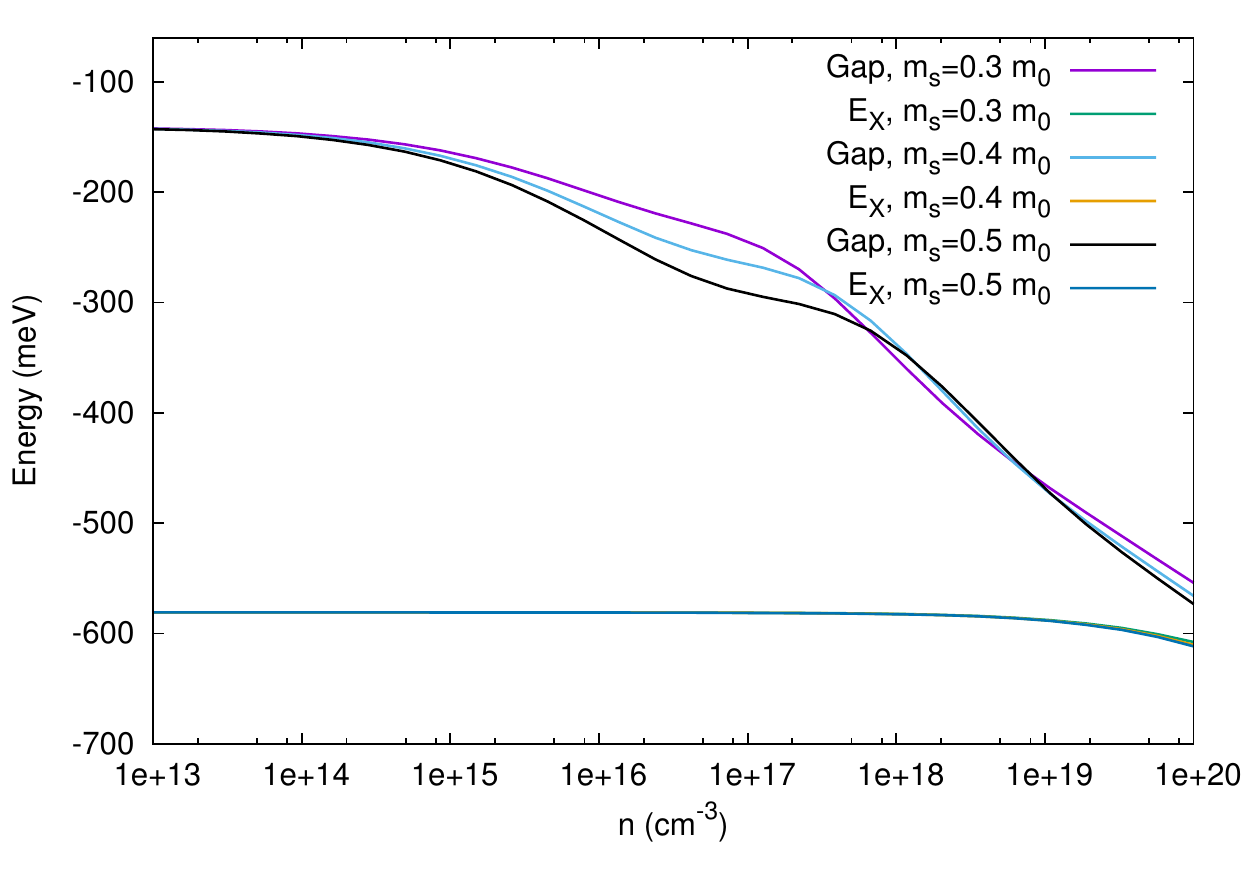}
\caption{Band-gap renormalization and 1s-exciton position relative to the unrenormalized band gap for MoS$_2$ on a substrate at $T=300$ K with $\varepsilon_{\infty}=2$ and quasi-particle broadening $\Gamma=10$ meV, shown for different carrier-doping densities and effective masses $m_{\textrm{s}}$.}
\label{fig:m_s_dep}
\end{figure}
\begin{figure}[h!t]
\centering
\includegraphics[width=\columnwidth]{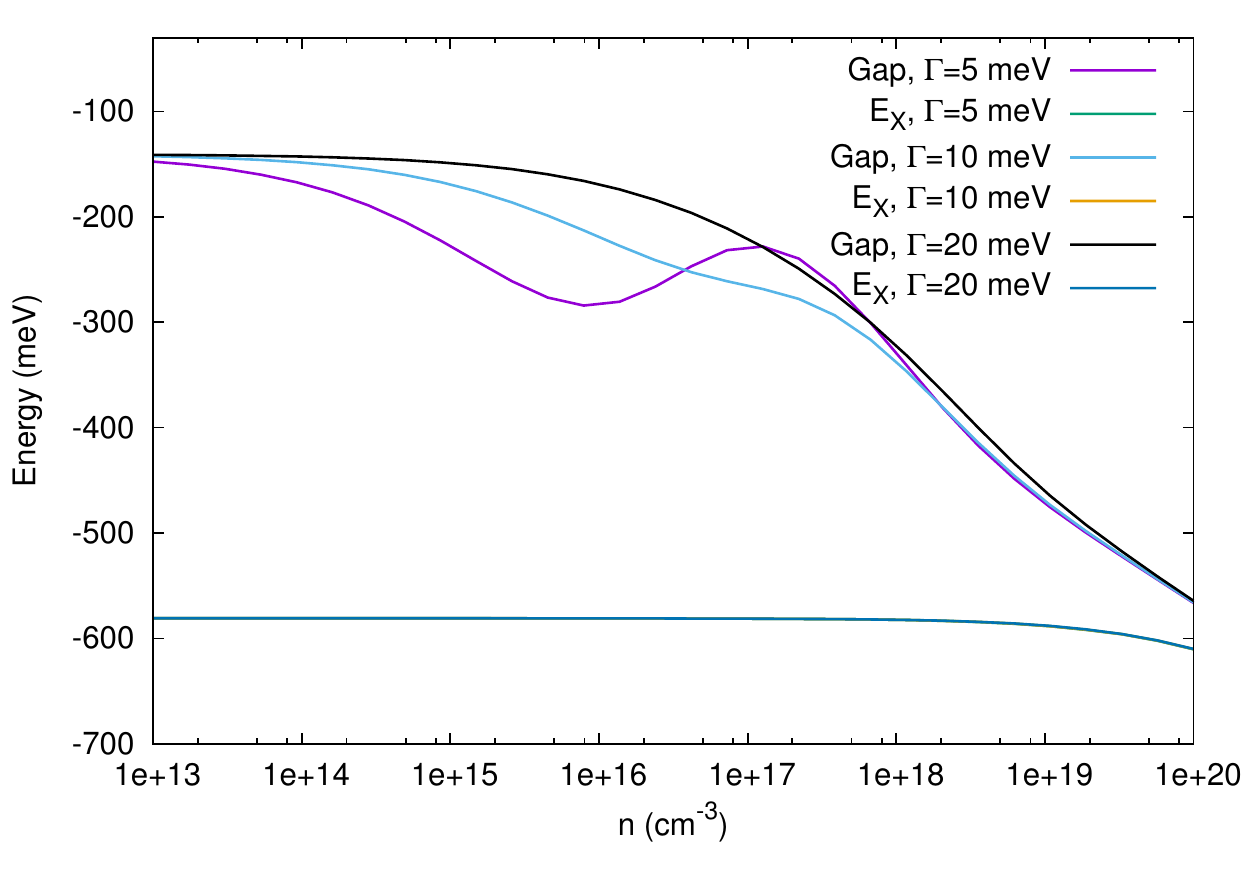}
\caption{Band-gap renormalization and 1s-exciton position relative to the unrenormalized band gap for MoS$_2$ on a substrate at $T=300$ K with $\varepsilon_{\infty}=2$ and $m_{\textrm{s}}=0.4\,m_0$, shown for different carrier-doping densities and quasi-particle broadenings $\Gamma$.}
\label{fig:gamma_dep}
\end{figure}

Here we discuss the dependence of the band-gap and exciton renormalizations on the model parameters $\varepsilon_{\infty}$, $m_{\textrm{s}}$ and $\Gamma$. The data is shown in Figs.\ref{fig:eps_inf_dep}, \ref{fig:m_s_dep} and \ref{fig:gamma_dep}, respectively. 
\par
With increasing high-frequency dielectric constant of the substrate due to inner electrons, $\varepsilon_{\infty}$, the static SXCH shift of the band gap increases as well. At the same time, the static screening of exciton binding energy due to inner electrons becomes more efficient, so that the exciton position exhibits only a small red shift. The dynamical shifts on the other hand become smaller due to the more efficient screening of exciton-plasmon interaction.  
\par
The effective mass $m_{\textrm{s}}$ of doped carriers in the substrate has a much smaller impact, modifying only the nontrivial part of the band-gap renormalization slightly. The exciton position is practically not affected.
\par
Finally, the quasi-particle broadening $\Gamma$ has a significant impact on the behaviour of the band gap at intermediate densities, as for weak broadening the renormalization becomes non-monotonous with density. The exciton on the other hand is not affected at all by the broadening. 

 
\bibliographystyle{apsrev}
 

%

\end{document}